\documentclass[reprint,aps,amsmath,amssymb,superscriptaddress,longbibliography,prx]{revtex4-2}
\usepackage{float,placeins,multirow,booktabs,todonotes,graphicx,dcolumn,bm,lipsum,layouts,url,siunitx}
\usepackage[colorlinks=true, linkcolor=byzantine, citecolor=emerald, urlcolor=emerald, breaklinks=true]{hyperref}
\usepackage{color
}
\usepackage{braket,physics}
\usepackage{soul}
\usepackage[d]{esvect}
\usepackage{nicefrac,tikz,circuitikz,amsmath,hyperref}
\usepackage{mathtools}
\usepackage[capitalise]{cleveref}

\makeatletter
\def\@bibdataout@aps{%
\immediate\write\@bibdataout{%
@CONTROL{%
apsrev41Control%
\longbibliography@sw{%
    ,author="08",editor="1",pages="1",title="0",year="1"%
    }{%
    ,author="08",editor="1",pages="1",title="",year="1"%
    }%
  }%
}%
\if@filesw \immediate \write \@auxout {\string \citation {apsrev41Control}}\fi 
}
\makeatother

\begin{document}
\definecolor{byzantine}{rgb}{0.74, 0.2, 0.64}
\definecolor{emerald}{rgb}{0.25,0.5,0.27}
\title{
Emergent macroscopic bistability induced by a single superconducting qubit 
}
\author{Riya Sett}
\email{riya.sett@ist.ac.at}
\affiliation{Institute of Science and Technology Austria (ISTA), 3400 Klosterneuburg, Austria}
\author{Farid Hassani}
\affiliation{Institute of Science and Technology Austria (ISTA), 3400 Klosterneuburg, Austria}
\author{Duc Phan}
\affiliation{Institute of Science and Technology Austria (ISTA), 3400 Klosterneuburg, Austria}
\author{Shabir Barzanjeh}
\affiliation{Institute of Science and Technology Austria (ISTA), 3400 Klosterneuburg, Austria}
\affiliation{Institute for Quantum Science and Technology, University of Calgary, Calgary, Alberta, Canada}
\author{Andras Vukics}
\email{vukics.andras@wigner.hu}
\affiliation{Wigner Research Centre for Physics, H-1525 Budapest, P.O. Box 49., Hungary}
\author{Johannes M.~Fink}
\email{jfink@ist.ac.at}
\affiliation{Institute of Science and Technology Austria (ISTA), 3400 Klosterneuburg, Austria}

\date{\today}

\begin{abstract}
The photon blockade breakdown in a continuously driven cavity QED system has been proposed as a prime example for a first-order driven-dissipative quantum phase transition. 
But the predicted scaling from a microscopic system - dominated by quantum fluctuations - to a macroscopic one - characterized by stable phases - and the associated exponents and phase diagram have not been observed so far. 
%
In this work we couple a single transmon qubit with a fixed coupling strength $g$ to an \textit{in-situ} bandwidth $\kappa$ tuneable superconducting cavity
to controllably approach this thermodynamic limit.
Even though the system remains microscopic, we observe its behavior to become more and more macroscopic as a function of $g/\kappa$. 
For the highest realized $g/\kappa\approx287$ the system switches with a characteristic dwell time as high as 6 seconds between a bright coherent state with $\approx 8\times10^3$ intra-cavity photons and the vacuum state with equal probability. This exceeds the microscopic time scales by six orders of magnitude and approaches the near perfect hysteresis expected between two macroscopic attractors 
in the thermodynamic limit. These findings and interpretation are qualitatively supported by semi-classical theory and large-scale 
Quantum-Jump Monte Carlo simulations. Besides shedding more light on driven-dissipative physics in the limit of strong light-matter coupling, this system might also find applications in quantum sensing and metrology.
%
\end{abstract}

\maketitle

\section{Introduction}
Quantum phase transitions (QPT), both first- and second-order \cite{vojta2003quantum} have been at the forefront of physics research for half a century. The original idea of QPTs as abrupt shifts in the (pure) \emph{ground state} of closed quantum systems as a function of a control parameter applied mostly to condensed matter physics. Dissipative phase transitions (DPT) occurring in the (in general, mixed) \emph{steady state} of open quantum systems \cite{Capriotti2005dissipation,Diehl2008,Nagy2010,Diehl2010,Kessler2012,LeBoite2013,Minganti2018,Hwang2018,Gutierrez2018,Reiter2020,Soriente2021}, however, broadened the scope of phase transitions to encompass mesoscopic and later even microscopic systems, where the interaction with the environment essentially affects the system dynamics. A DPT was first realized experimentally in a Bose-Einstein condensate interacting with a single-mode optical cavity field \cite{Baumann2010}, and DPTs are increasingly relevant to today's quantum science and technology \cite{verstraete2009quantum,fernandez2017quantum,Fitzpatrick2017,garbe2020critical}.

In view of this success, it is remarkable that in recent years yet another phase-transition paradigm could emerge, namely, \emph{first-order} dissipative quantum phase transitions. A first-order phase transition means that two phases can coexist in a certain parameter region, like water and ice at $\SI{0}{\celsius}$ for a certain range of free energy. Coexistence of phases in the quantum steady state seems paradoxical, since the steady-state plus normalization conditions for the density operator constitute a linear system of equations, that admits only a single solution. That is, given the Liouvillian superoperator $\mathcal L$ for the Markovian evolution of the system, there exists only a single normalized density operator $\rho_\text{st}$ that satisfies
\begin{equation}
 \mathcal L \rho_\text{st}=0.
\end{equation}
The resolution is that a single density operator can accommodate the mixture of two macroscopically distinct phases expressed as a ratio of the two components. In the water analogy, at $\SI{0}{\celsius}$ we could symbolically write
\begin{equation}
\label{eq:waterIce}
 \rho_\text{st}=c\,\rho_\text{water}+(1-c)\,\rho_\text{ice},
\end{equation}
with $c$ growing from 0 to 1 as the free energy is increased.

Recently, first-order dissipative quantum phase transitions have been found in various systems. One such platform is the clustering of Rydberg atoms described by Ising-type spin models \cite{Ates2012,marcuzzi2014universal,Overbeck2017,Roscher2018,samajdar2021quantum,myerson2022construction} and realized experimentally \cite{Carr2013,Malossi2014,Letscher2017}. Various other systems of ultracold atoms \cite{Labouvie2016,Ferri2021} and dissipative Dicke-like models \cite{Gelhausen2018,Stitely2020} also exhibit signatures of a first-order DPT. Other platforms include (arrays of) nonlinear photonic or polaritonic modes \cite{LeBoite2013,Casteels2017,debnath2017nonequilibrium,Rodriguez2017,Savona2017,TFink2018,Vicentini2018,Lang2020,Li2022}, exciton-polariton condensates \cite{Hanai2019,Dagvadorj2021} and circuit QED \cite{Mavrogordatos2017,Fitzpatrick2017,Fink2017,Brookes2021}. In this work we observe and model the scaling and phase diagram of a first-order DPT in zero dimensions, i.e. for a single qubit strongly coupled to a single cavity mode. 


\section{Photon-Blockade breakdown}
The Jaynes-Cummings (JC) model - one of the most important models in quantum science - describes the  interaction between atoms and photons trapped in a cavity \cite{Haroche2006}. It is expressed by the Hamiltonian ($\hbar=1$)
\begin{multline}
\label{eq:JaynesCummings}
H_\text{JC}=\omega_\text{R}\,a^\dagger a\,+\,\omega_\text{A}\, \sigma^\dagger\,\sigma\, +\, i g \qty( a^\dagger\,\sigma-\sigma^\dagger\,a )\\+\,i \eta\qty( a^\dagger\,e^{-i\omega t}-a\,e^{i\omega t})\; ,
\end{multline}
with $\omega_\text{R}$ the angular frequency of the cavity mode with boson operator $a$, $\omega_\text{A}$ that of the atomic transition with operator $\sigma$, $g$ the coupling strength, $\eta$ the drive strength, and $\omega$ the drive frequency. 
This model yields the prototype of an anharmonic spectrum in the strong-coupling regime, as demonstrated in cavity \cite{brune1996quantum} and circuit QED \cite{fink2008climbing}, and with quantum dots in semiconductor microcavities \cite{kasprzak2010up}. Its strong anharmonicity at single photon levels is the basis of the photon blockade effect \cite{Imamoglu1997,Lang2011}, in analogy with Coulomb blockade in quantum dots or to polariton blockade \cite{ohira2021polariton}. Photon blockade means that an excitation cannot enter the JC system from a drive tuned in resonance with the bare resonator frequency, or similarly, a second excitation from a drive tuned to resonance with one of the single-excitation levels cannot enter. 

This blockade is, however, not absolute, as it can be broken \cite{Carmichael2015,Dombi2015,Curtis2021,Palyi2012} by strong enough driving due to a combination of multi-photon events and photon-number increasing quantum jumps \cite{Vukics2019}. In an intermediary $\eta$ range, in the time domain the system stochastically alternates between a blockaded, \emph{dim state} without cavity photons and a \emph{bright state} in which the blockade is broken and the system resides in the highly excited quasi-harmonic part of the spectrum resulting in a large transmission of drive photons. In phase space, this behavior results in a bimodal steady-state distribution 
\begin{equation}
\label{eq:brightDim}
 \rho_\text{st}=c\,\rho_\text{bright}+(1-c)\,\rho_\text{dim},
\end{equation}
in analogy with \cref{eq:waterIce}, with $c$ growing from 0 to 1 with increasing $\eta$. This effect has been demonstrated experimentally in a circuit QED system \cite{Fink2017}.

Bistability in the time domain or bimodality in phase space is, however, not sufficient evidence for a first-order phase transition. It is also necessary that the two constituents in the mixture \cref{eq:brightDim} corresponding to the two states in the temporal bistable signal to be macroscopically distinct as is the case in \cref{eq:waterIce}. It has been shown theoretically \cite{Carmichael2015,Vukics2019}, that the photon blockade breakdown (PBB) effect has such a regime, i.e. a \emph{thermodynamic limit}, where both the timescale and the amplitude of the bistable signal goes to infinity, resulting in long-lived and macroscopic distinct dim and bright phases. Remarkably, this thermodynamic limit is a strong-coupling limit, defined as $g\to\infty$, and independent of the physical system size, i.e. the system remains the same JC system composed of two microscopic interacting subsystems. In this limit, the temporal bistability is replaced by hysteresis, where the state of the system is determined by its initial condition, since switching to the other state entails an infinite waiting time. The passage to the thermodynamic limit, i.e. the indefinite increase of $g$ has been termed \textit{finite-size scaling} \cite{Vukics2019}. 

In this work, we demonstrate these additional criteria that clearly signify the observed physical effect as a first-order dissipative quantum phase transition. We demonstrate the finite-size scaling over 7 orders of magnitude towards the thermodynamic limit and back out the phase diagram of a first-oder DPT in zero dimensions. We realize this experiment with a superconducting qubit strongly coupled to a bandwidth-tunable microwave cavity mode and find qualitative agreement with large-scale Quantum-Jump Monte Carlo simulations and semi-classical calculations of the phase boundaries.

\section{Experimental Realization}
\label{sec:system}
\begin{figure}
\centering
\includegraphics[width=\columnwidth]{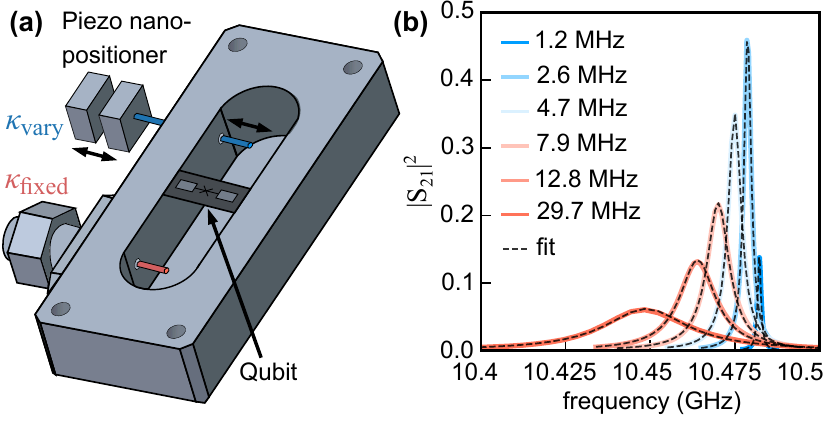}
\caption{\textbf{Experimental realization. (a)} Schematics of the experimental device consisting of a superconducting transmon qubit fabricated on a silicon substrate that is placed at the antinode of the fundamental mode of a 3D copper cavity. The cavity has a fixed length port (red) and an in-situ variable length pin coupler port (blue). 
\textbf{(b)} Measured cavity transmission spectra with the qubit far detuned for different coupler positions (color coded) together with a fit to \cref{eq:ipop} (dashed) and the extracted $\kappa/2\pi$.  
}
\label{fig:setup}
\end{figure}

Our experimental setting incorporates a transmon qubit \cite{Koch2007, Paik2011} placed at the anti-node of the standing wave of a 3D copper-cavity, as shown in \cref{fig:setup}(a), that can be flux-tuned by applying a B-field via a millimeter-sized superconducting bias coil mounted at the outside cavity wall. The transmon qubit has a maximum Josephson energy $E_\text{J,max}/h \approx \SI{48}{\giga\hertz}$, charging energy $E_\text{C}/h \approx \SI{382}{\mega\hertz}$ and a 
resulting maximum transition frequency between its ground and first excited states of $\omega_\text{A}/2\pi \approx\SI{12.166}{\giga\hertz}$.
When the transmon ground to first excited state transition is tuned in resonance with the cavity mode at $\omega_\text{R}/2\pi \approx\SI{10.4725}{\giga\hertz}$, the directly measured coupling strength between the single photon and the qubit transition is as high as $g/2\pi = \SI{344}{\mega\hertz}$, which is only about a factor of 3 below the so-called ultrastrong coupling regime \cite{forn2019ultrastrong}. The relatively high absolute anharmonicity between subsequent transmon state transitions is 
$\alpha/h \approx -\SI{418}{\mega\hertz}$
at this flux bias position. 

The cavity has two ports, of which the input pin coupler position is fixed with an external coupling strength of $\kappa_\text{fixed}/2\pi\approx\SI{500}{\kilo\hertz}$.
The output coupler is attached to a cryogenic piezo nano-positioner, which allows for adjusting the pin length extending into the cavity \cite{Sears2012}. With this tunable coupler the coupling strength can be varied in situ in a wide range $\kappa_\text{vary}/2\pi\approx\SI{20}{\kilo\hertz} - \SI{30}{\mega\hertz}$.
The 
internal cavity loss at low temperature is
$\kappa_\text{int}/2\pi\approx\SI{600}{\kilo\hertz}$, which is achieved by electro-polishing of the high conductivity copper surface before cooldown to 10~mK in a dilution refrigerator.
 
All four scattering parameters are measured with a vector network analyzer to calibrate the measurement setup and the cavity properties when the qubit is far detuned from the cavity resonance. Figure \ref{fig:setup}(b) shows transmission measurements fitted with the scattering parameter $S_\text{21}$ 
derived from the Input-Output theory of an open quantum system \cite{Gardiner1985} 
\begin{equation}
\label{eq:ipop}
S_\text{21} = \frac{\sqrt{\kappa_\text{fixed} \kappa_\text{vary}}}{\kappa/2 - i(\omega - \omega_\text{R})}.
\end{equation}
From these fits, we extract all loss rates that add up to the total cavity linewidth $\kappa= \kappa_\text{fixed} + \kappa_\text{vary} + \kappa_\text{int}$ also indicated in \cref{fig:setup}(b). 

Time-domain characterization measurements confirm that the qubit is Purcell-limited and homogeneously broadened at the flux sweet spot \cite{Houck2008}, where the measured coherence times are $T_1 \approx \SI{0.5}{\micro\second}$ and $T_2 \approx\SI{1}{\micro\second}$. When the qubit frequency is tuned far below the resonator frequency $\omega_\text{A}/2\pi \approx\SI{6.083}{\giga\hertz}$ by applying an external magnetic field, the measured coherence times are $T_1 \approx \SI{18.14}{\micro\second}$ and $T_2 \approx \SI{0.496}{\micro\second}$, which we attribute to a higher Purcell limit due to the larger detuning as well as a drastically increased flux noise sensitivity. On resonance $\omega_\text{A}=\omega_\text{R}$, where the following experiments were performed, the energy relaxation is therefore fully dominated by cavity losses. The measured vacuum Rabi peak linewidth changes with and without the qubit in resonance are in agreement with a small amount of flux noise induced dephasing expected at that flux bias position. 


\section{photon blockade breakdown measurement}
\label{sec:PBB}

\begin{figure*}
\centering
\includegraphics[width=0.99\linewidth]{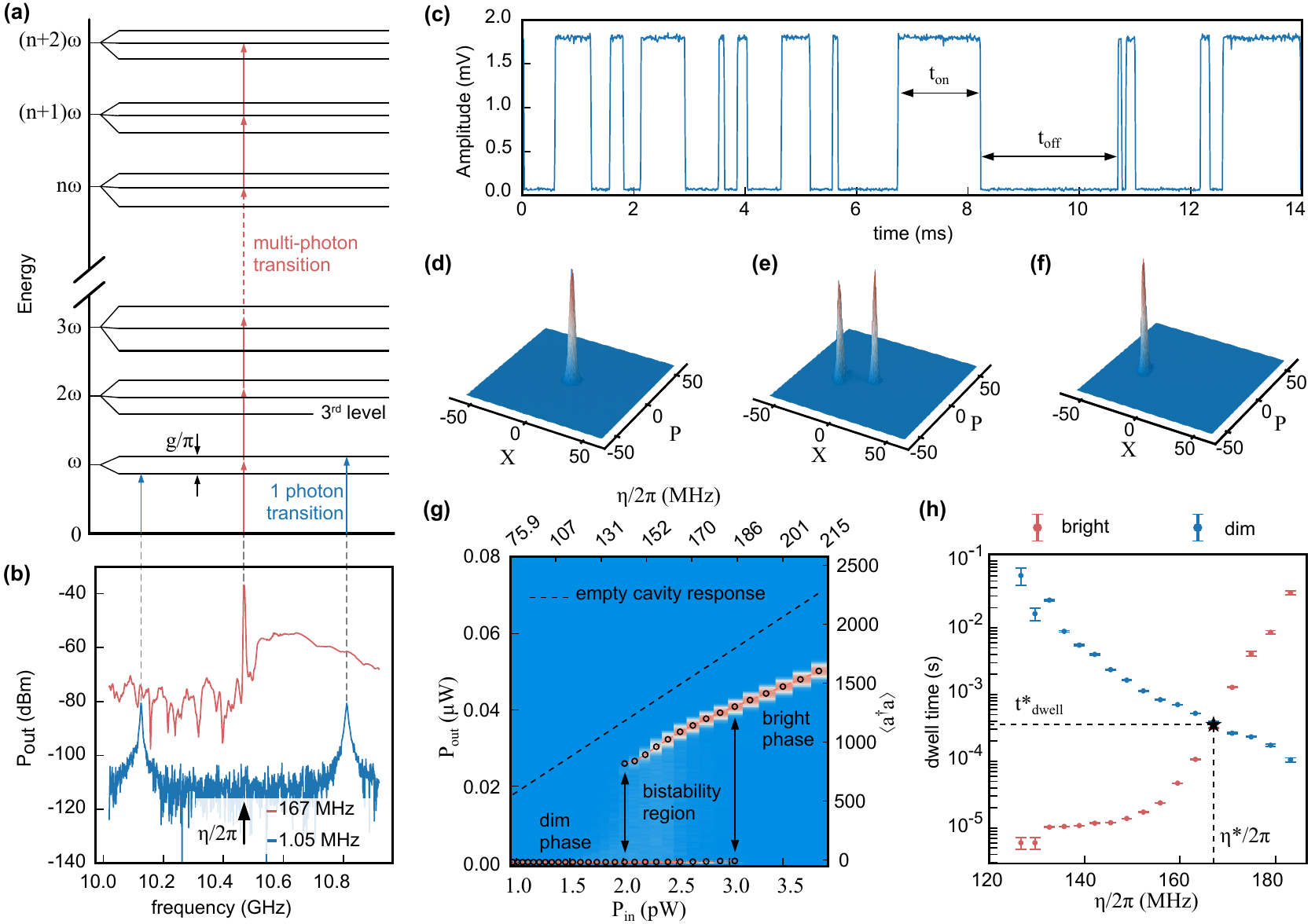}
\caption{\textbf{Observation of photon blockade breakdown at $g/\kappa\approx 39$.}
\textbf{(a)} The Jaynes-Cummings ladder for a three-level atom illustrating the PBB effect in the frequency domain: single-photon (blue) and multi-photon transitions (red) are indicated according to the measured spectrum at the Rabi-split frequencies and near resonance, respectively. 
\textbf{(b)} Measured cavity transmission spectra for $\omega_\text{A}=\omega_\text{R}$ for two applied external drive strengths $\SI{1.05}{\mega\hertz}$ revealing a typical vacuum Rabi spectrum and $\SI{167.01}{\mega\hertz}$ where a sharp peak at $\omega_\text{R}$ is observed.
\textbf{(c)} Measured cavity output bistability at $\omega_\text{R}$ in the time domain indicating the dwell times of the bright ($t_\text{on}$) and dim states ($t_\text{off}$).
\textbf{(d-f)} Measured quadrature histograms (proportional to cavity Q-functions convolved with amplifier noise) for the dim phase at $\eta/2\pi=\SI{105}{\mega\hertz}$, for the bistable region with equal probability at $\SI{167}{\mega\hertz}$, and for the bright phase at $\SI{210}{\mega\hertz}$, respectively. 
\textbf{(g)} Measured histograms of the output power (arranged vertically and probability is color coded) as a function of input power.
The maxima indicated by circles trace out a typical bistability curve, cf. \cref{fig:bistability} in \cref{app:semiclassical}. 
\textbf{(h)} Extracted average dwell times in the dim and bright state (\cref{eq:times}) as a function of $\eta$. 
The error bars represent the standard error that is extracted from 5 sections of the full data set. The dwell-time and drive strength corresponding to half-filling where $t_\text{dwell}^\text{dim}=t_\text{dwell}^\text{bright}$ are indicated with an asterisk.}
\label{fig:pbb}
\end{figure*}

The photon blockade (and its breakdown) phenomenon most straightforwardly occurs when the two interacting constituents are resonant $\omega_\text{A}=\omega_\text{R}$. In contrast to the ideal two-level atom limit \cite{Carmichael2015,Vukics2019}, when driven on resonance $\omega=\omega_\text{R}$ this does not lead to spontaneous dressed-state polarization \cite{alsing1991spontaneous,armen2009spontaneous} - a second-order DPT \cite{Carmichael2015}, in our experimental situation with three (or more) transmon levels \cite{Fink2017} as shown in \cref{fig:pbb}(a).
For low input powers corresponding to less than a single intra-cavity photon on average we observe a vacuum Rabi-split spectrum in transmission, as shown in \cref{fig:pbb}(a, b) (blue line). 
No transmission peak is observed at the bare cavity frequency $\omega_\text{R}$ 
up to intermediate input drive strengths $\eta$. This means that a single photon - or even hundreds of photons at the chosen $g/\kappa = 39.1$ - are prevented from entering the cavity due to the presence of a single artificial atom. 

This blockade is observed to be broken abruptly by further increasing the applied drive strength $\eta$, which is proportional to square-root of the applied drive power and the corresponding drive photon number. As $\eta$ is increased by only a finite amount, the transmitted output power increases by three orders of magnitude at the bare resonator frequency, as shown in the red spectrum in \cref{fig:pbb}(b). The central sharp peak in the transmission spectrum corresponds to a time-averaged measurement (determined by the chosen resolution bandwidth) of a cavity that is fully transparent for most of the integration time. 
This PBB effect can be attributed to the nonlinearity of the lower part of the JC spectrum which is strongly anharmonic \cite{fink2008climbing,Bishop2009}, while the 
higher-lying part of the spectrum has subsets 
that are closely harmonic over a certain range of excitation numbers \cite{Fink2010} and can hence accommodate a closely coherent state. 

In the time domain, with $\eta$ in the phase coexistence region, the PBB effect results in a bistable telegraph signal, where the system output alternates between a `dim' state where the qubit-resonator system remains close to the vacuum state unable to absorb an excitation from the externally applied drive, and a `bright' state where the system resides in an upper-lying, closely harmonic subset of the JC spectrum, cf. \cref{fig:pbb}(c). The switches between these two classical attractors are necessarily multi-photon events that are triggered by quantum fluctuations. This bistability was shown to be a finite-size precursor of what would be a first-order DPT in the thermodynamic limit ($g/\kappa\to\infty$) \cite{Vukics2019}, where the bistability develops into perfect hysteresis: the system is stuck in the attractor determined by the initial condition as long as the control parameters are set 
in the transition domain.

In order to investigate this dynamics qualitatively, we record the real-time single-shot data of both quadratures of the transmitted output field at the bare cavity frequency while applying a continuous-wave (CW) drive tone resonant with the bare cavity, over a range of applied drive strengths. 
The transmitted radiation is first amplified with a high electron mobility transistor (HEMT) at $\SI{4}{\kelvin}$ followed by another room-temperature low-noise amplifier (LNA), then down-converted with an IQ mixer with appropriate IF frequency and finally digitized with a digitizer. 
Further this recorded data is digitally low-pass filtered with appropriate resolution bandwidth and down-converted to d.c. 
to extract the time-dependent quadratures 
in voltage units. For example, in the case of $\kappa/2\pi=\SI{8}{\mega\hertz}$, the recorded data is $\SI{2.88}\second$ long and the final time resolution of the extracted quadratures is $\SI{2.5}{\micro\second}$, cf. \cref{fig:pbb}(c). The selection of an appropriate resolution bandwidth is critical for a number of reasons: (1) to successfully resolve frequent and sudden switching events caused by very short dwell times at high $\kappa$ values, (2) to maintain a signal to noise ratio that allows to clearly discriminate single shot measurement events without averaging, and (3) to achieve a sufficient total measurement time to resolve long dwell times with the available memory. 

From the resulting histograms in phase space, cf.~\cref{fig:pbb}(d-f), which represent the scaled Husimi-Q functions convolved with the added amplification chain noise photon number $n_\text{amp}\approx9.2$, it can be deduced that for low drive strength the photon blockade is intact (dim phase) with the Q function being centered around the vacuum state. Upon increasing the input drive strength, the Q function becomes bimodal with decreasing weight of the dim state as described in Eq.~\ref{eq:brightDim}. At high enough $\eta$ only the bright coherent state is measured. Note that the transformation of the Q-function and hence the steady-state density operator of the system as a function of $\eta$ is continuous, yet a first order phase transition with a well-defined coexistence region can occur. 

A similar conclusion can be drawn from the output power histograms (color map) that trace out a typical bistability curve as shown in \cref{fig:pbb}(g). The most likely output powers $P_\text{out}$ and calculated equivalent intra-cavity photon numbers 
of the empty cavity driven on resonance $\bar{n}_\text{cav}=\frac{P_\text{in}}{\hbar\omega_\text{R}}\frac{4\kappa_\text{fixed}}{\kappa^2}$ as a function of applied input power $P_\text{in}$ and resulting drive strength $\eta=\sqrt{\bar{n}_\text{cav}}\kappa/2$ are marked with circles.  The vacuum, bistability and bright regions are well defined. We find that the derivative of the bright solution obtained at high $P_\text{in}$ deviates somewhat from the empty cavity response measured when the qubit is far detuned (dashed line). For large $g/\kappa$ values this is more pronounced and we have observed that this can lead to secondary bistability regions at even higher powers for $g/\kappa \gtrsim 43$, 
which we believe to originate from the multi-level nature of the transmon qubit. In this work we focus only on the  bistability occurring at the lowest drive strength. 

In \cref{fig:pbb}(h) we show the measured dwell times of dim and bright states as a function of input power. The average dwell times at each input power are calculated as
\begin{equation}
\label{eq:times}
\begin{aligned}
t_\text{dwell}^\text{dim} &= \frac{1}{N} \sum_{n=1}^{N} t_{\text{off},n} ; 
\;\;\;\; 
t_\text{dwell}^\text{bright} &= \frac{1}{N} \sum_{n=1}^{N} t_{\text{on},n} ;
\end{aligned}
\end{equation}
and the threshold for one of the $N$ switching events during the full measurement duration (with 2.5\,$\mu$s resolution) is defined at half of the observed full amplitude for the lowest applied $\eta/(2\pi) \approx 140$\,MHz where the bistability is fully developed. Note that in cases of low signal to noise ratio, e.g. for large $\kappa$ or at high drive detunings shown later, we used a higher threshold that at least exceeds the variance of the output power of the dim state.

At low $\eta$ the statistics is low because the system remains in the dim state for long time scales. As $\eta$ is increased to $\eta^*/(2\pi)\approx167$\,MHz the measured average dwell times cross at $t_\text{dwell}^\text{dim}=t_\text{dwell}^\text{bright}=t^*_\text{dwell}\approx354\,\mu$s. We call this the half-filling point where it is equally likely for the system to be found in the dim or bright state, denoted by an asterisk in \cref{fig:pbb}(h). For even higher drive strength the system prefers to dwell in the bright state, i.e. we observe close to full resonator transmission for most of the time. 

\section{Finite size scaling towards the thermodynamic limit}
\label{sec:TL}
We set out to experimentally prove the scaling of the measured PBB bistability towards a thermodynamic limit, i.e. to show - despite the underlying microscopic nature of the system - a truly macroscopic behavior as $g/\kappa \rightarrow \infty$ where it has been shown to become a first-order DPT. We also experimentally determine the finite-size scaling exponents of the characteristic time and the corresponding drive strength and intra-cavity photon number for this DPT.

\begin{figure*}
\centering
\includegraphics[width=0.99\textwidth]{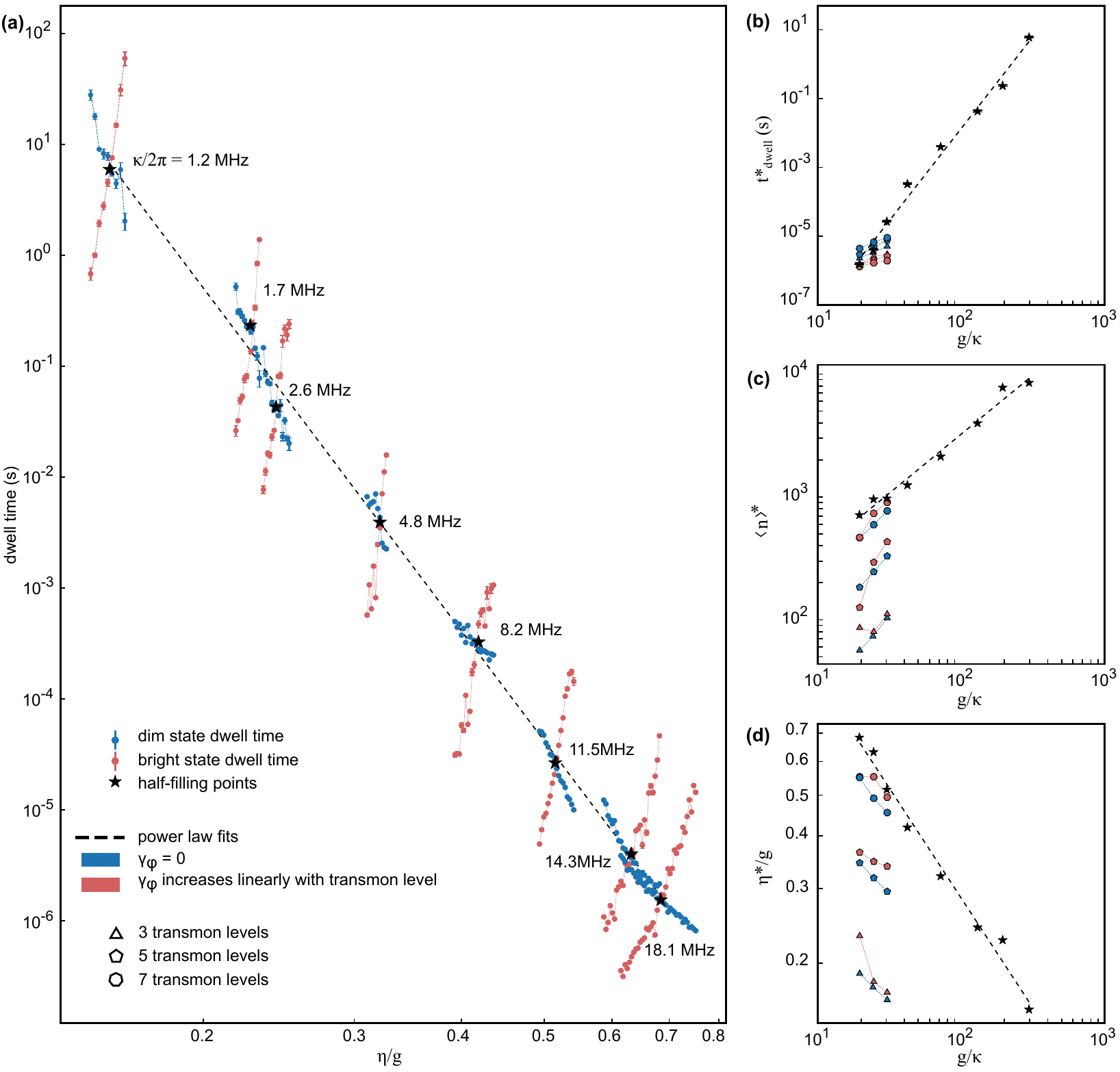}
\caption{\textbf{Finite size scaling towards the thermodynamic limit.} 
\textbf{(a)} 
Measured dwell times in the dim (blue) and bright states (red) are shown as a function of $\eta/g$ for different values of $\kappa$. Black stars denote half-filling values where the probabilities to be in the dim or bright state are equal. The dashed line is a power law fit. 
%
The error bars represent the standard error that is extracted from at least 3 sections of the full data set.
\textbf{(b)-(d)} Measured scaling laws as a function of control parameter $g/\kappa$ 
at half filling (asterisks). 
Fitting the experimental data (dashed lines) yields the exponents $t_\text{dwell}^* \propto(g/\kappa)^{5.4\pm0.2}$ for the dwell time, $\langle n\rangle^* \propto(g/\kappa)^{0.96\pm0.05}$ for the average resonator photon number and $\eta^*/g \propto (g/\kappa)^{-0.52\pm0.03}$ for the input drive strength. We compare the experimental values (stars) with those obtained from QJMC simulations (polygons with color code) for computationally manageable $\kappa=11.5,\;14.3,\;\text{and}\;18.1~\text{MHz}$. The simulations include 3, 5, or 7 transmon levels both, without transmon dephasing (light blue) and with dephasing increasing linearly with the level number (light red) with $\gamma_{\phi,1}/(2\pi)=50\,$kHz and $\gamma_1=0$. Details of the simulation are in the main text and in \cref{app:fullQuantum}. 
} 
\label{fig:thermodynamic_limit}
\end{figure*}

In \cref{fig:thermodynamic_limit}(a) we show the measured dwell times as a function of drive strength - similar to \cref{fig:pbb}(h) - for different  $\kappa/2\pi$ ranging from $\SI{18.1}{\mega\hertz}$ to $\SI{1.2}{\mega\hertz}$.
For each $\kappa$ value the dwell time in the dim state (blue symbols) decreases with increasing drive strength and that of bright state (red symbols) increases until eventually the system is fully stabilized in the bright state. For each $g/\kappa$ value, we define a single characteristic time of the process $t^*_\text{dwell}$ at the drive strength $\eta^*$ that leads to half-filling of the telegraph signal, cf. also \cref{fig:pbb}(h). 

Remarkably, for the largest realized value of $g/\kappa=287$, the characteristic dwell time reaches $t^*_\text{dwell}\approx\SI6\second$. This exceeds the characteristic microscopic dissipation times of order $\kappa^{-1}$ by a factor $\sim10^6$ and is reminiscent of the emergence of two macroscopically distinct states 
 with strongly suppressed transitions that require a 
 cascade of quantum jumps 
 \cite{Vukics2019}. This strong coupling, high photon number limit where the effect of quantum fluctuations vanishes has been defined as the thermodynamic limit of such a finite-size zero-dimensional system \cite{Carmichael2015}.

Finite-size scaling towards the thermodynamic limit means that
the characteristic time scales and brightness scale up as a function of $g/\kappa$, while the system remains self-similar, which in this case means that it keeps switching stochastically at a fixed filling factor (that we choose 0.5). 
In \cref{fig:thermodynamic_limit}(b) we plot the measured increase of $t^*_\text{dwell}$ at filling 0.5 
over a range of seven orders of magnitude as a function of $g/\kappa$. The behavior follows a strong power law over the full range and the fitted finite-size scaling exponent is
$t^*_\text{dwell} \propto (g/\kappa)^{5.4\pm0.2}$.
Similarly, the intra-cavity photon number of the bright state at half-filling increases nearly linearly with 
$\expval{n}^* \propto (g/\kappa)^{0.96\pm0.05}$, as shown in \cref{fig:thermodynamic_limit}(c), 
and as a consequence the corresponding drive strength decreases with the square root
$\eta^*/g \propto (g/\kappa)^{-0.52\pm0.03}$. Here the drive normalization with $g$ is motivated by the 2-level neoclassical theory, where the critical point appears at $\eta/g=0.5$ for $\Delta=0$.

The theoretical results (triangles, pentagons and heptagons) shown in \cref{fig:thermodynamic_limit}(b)-(d) are taken from large scale numerical simulations performed with C++QED: a framework for simulating open quantum dynamics \cite{Vukics2012}. 
An adaptive version of the Quantum-Jump Monte Carlo (QJMC) method is applied, where a single stochastic quantum trajectory is considered to correspond to a single experimental run \cite{kornyik2019monte}. The shown data is based on 64 CPU years of simulation time with a Hilbert space dimension of up to $\sim 2^{14} - 2^{15}$ (7 transmon levels and 3-5 times $\expval{n}^*$). Another computationally demanding aspect is that the required time step is set by the largest characteristic frequency (typically $g$ or $\eta$) of the microscopic system (sampled with $1/\kappa$ to reduce data volume) whereas the total trajectory needs to cover many times $t_\text{dwell}$ to obtain sufficient statistics of the macroscopic behaviour. Together this limits the range of numerically accessible $g/\kappa$ values to the lowest three values investigated.
For more details on how we model the system, examples of simulated quantum trajectories and the impact of different transmon dephasing models, cf. \cref{app:fullQuantum}. 


The observed photon number scaling exponent is about half of the analytical prediction of $\expval{n}^*\propto (g/\kappa)^2$. This is not surprising since the 2-level neoclassical theory does not yield quantitative agreement for the case with a multi-level transmon circuit. Numerical simulations for the lowest three $g/\kappa$ values taking into account up to 7 transmon levels (heptagons in \cref{fig:thermodynamic_limit}(c)), agree with the measured linear exponent to within 15\%. The absolute value of $\expval{n}^*$ also agrees well (blue) and is further improved when we include qubit dephasing of all transmon levels (red) due to flux noise with the measured $\gamma_\phi/2\pi\approx 50$~kHz for the lowest qubit transition. That is, as long as enough transmon levels are taken into account. In the case of just 3 transmon levels the simulated value is about an order of magnitude smaller than the measured one. This highlights the importance and participation of multiple transmon levels in the dynamics of the system. 


The dwell time values and scaling shown in \cref{fig:thermodynamic_limit}(c) are more robust with regards to the number of transmon levels but we observe a substantial deviation between the measured (5.4) and simulated scaling exponents in the range of 0.9-2.2. In \cite{Vukics2019} based on a two-level model with finite detuning, the blink-off rate could be calculated from the rate of ladder-switching quantum jumps, and was found to be proportional to $\kappa/\langle n\rangle^*$, so the waiting time for a blink-off is $\langle n\rangle^*/\kappa$, therefore it scales as $g^2/\kappa^3$. The numerically determined timescale-exponent in the same work was $(\kappa t_\text{dwell}^*) \propto (g/\kappa)^{2.2}$, which is very close to the analytical value. 

The observed scaling exponent is significantly larger which could be due to counter-rotating terms not taken into account in the numerical simulations and due to hybridized decay channels which can invalidate our approach of using separate transmon and resonator decay channels in the master equation.
Other potentially participating mechanisms could include transmon ionization \cite{Shillito2022} or dielectric surface loss saturation \cite{Zemlicka2022} that might further stabilize the system - in particular in the bright attractor and thus reduce the blink-off probability. Taken together these effects appear to be driving the system significantly faster to the thermodynamic limit compared to what would be expected from the standard Jaynes-Cummings model. 

Importantly and irrespective of the origin of the unexpectedly strong scaling, we observe that the system is always able to relax to the vacuum state eventually - despite the continuous driving. And this vacuum state is then stabilized for seconds by the presence of a single qubit, even for the highest drive strength corresponding to a photon number of $\expval{n}\approx 10^4$. In contrast to similarly looking fluorescence signals with dwell times on the order of seconds, which have been known in quantum science since the famous first electron-shelving experiments with single-ions in Penning-traps \cite{Nagourney1986Shelving,Bergquist1986Observation} due to long lived metastable atomic states, in the present case the measured time scale exceeds all microscopic time scales by up to a factor $10^6$, i.e. the system is very deep in the macroscopic limit which justifies its classification as a finite-size phase transition. 

\section{Phase Diagram}
\label{sec:PD}

\begin{figure*}[t]
\centering
\includegraphics[width=0.99\textwidth]{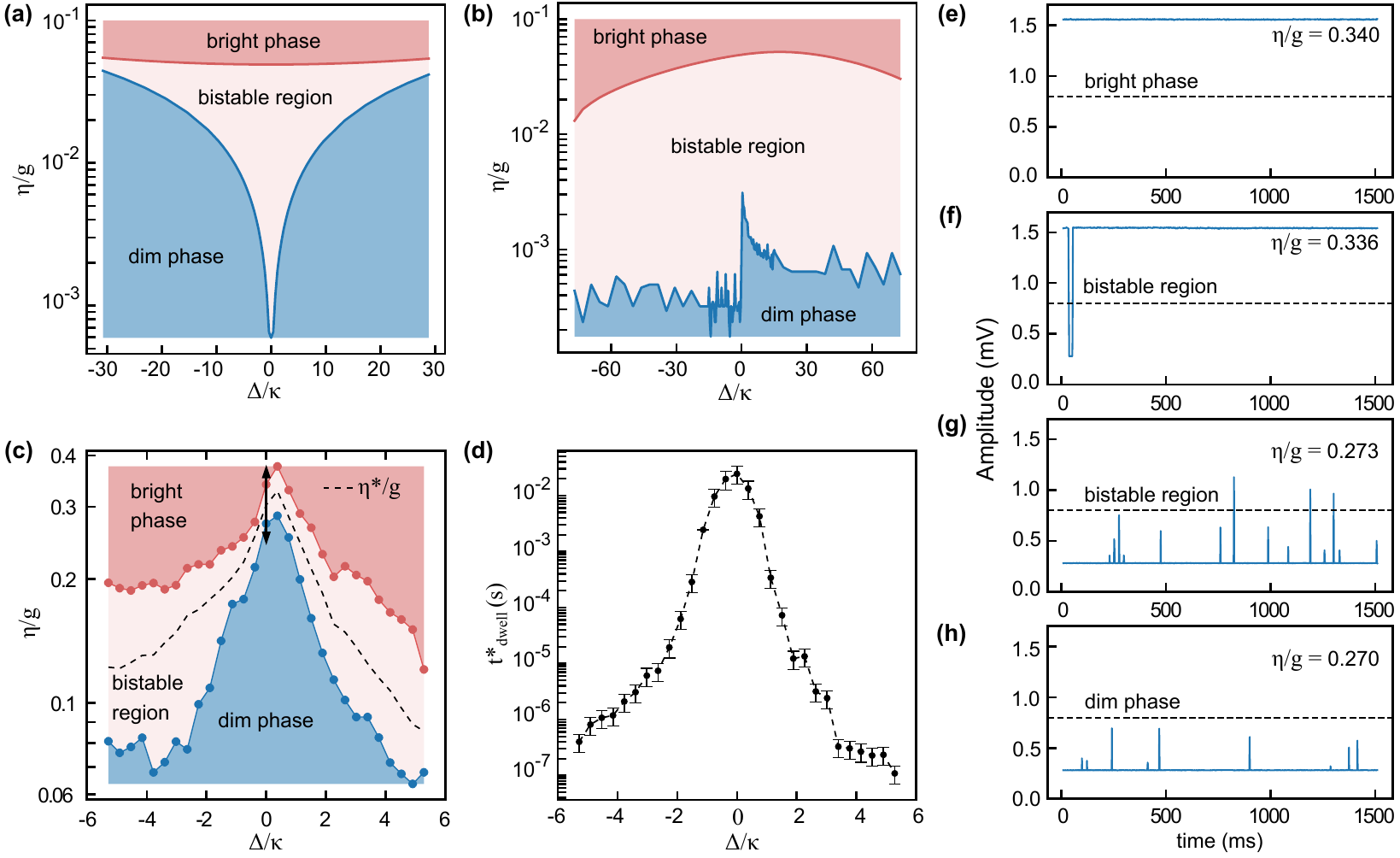}
\caption{\textbf{PBB phase diagram at $g/\kappa\approx 132$.} 
Phase diagram on the $\Delta-\eta$ plane obtained from semiclassical Maxwell-Bloch equations for \textbf{(a)} two and \textbf{(b)} three transmon levels with values for the parameters $\gamma_1$ and $\gamma_{\phi}$ according to the experiment. In panel (b), the noisy lower boundary of the bistable region is a result of numerical errors. \textbf{(c)} The PBB phase diagram with boundaries obtained from the experimental data (points) and the half-filling drive strength (dashed line). \textbf{(d)} Measured dwell times at half filling as a function of drive detuning. Error bars are extracted from the mean and standard error of the measured dim and bright dwell times.
\textbf{(e)-(h)} show the experimentally observed telegraph signals that define the phase boundaries 
at $\Delta = 0$ in the range of $\eta/2\pi$ from 85.6 to 115.5\,MHz as indicated by a double arrow in panel (c).
}
\label{fig:phase_diagram}
\end{figure*}

The PBB phase transition has been predicted to give rise to an interesting phase diagram as a function of drive detuning $\Delta \equiv \omega - \omega_\text{R} = \omega - \omega_\text{A}$. From the neoclassical equations, valid for pure two level qubit states and $\gamma_1=\gamma_\phi=0$, the bistable region is only expected at finite detunings away from the critical point at $\Delta=0$ \cite{Carmichael2015,Vukics2019}. A more realistic prediction is obtained from numerical solutions of the semiclassical Maxwell Bloch equations that include qubit decay
(for details, cf. \cref{app:semiclassical}), which are shown in \cref{fig:phase_diagram}(a),(b) for a 2- and a 3-level transmon, respectively. The results are qualitatively different not only in comparison to the neoclassical solution but also between the cases with 2 and 3 level transmons. Most importantly, in the case of 3 levels the bright phase boundary forms a peak rather than a dip at zero detuning and the dim phase is also predicted to exists around zero detuning.  

Experimentally we choose a large $g/\kappa\approx132.3$, where the time scales are long and the phases are very well defined, to back out the phase diagram as a function of $\Delta$. We sweep the drive strength for each chosen detuning and record a trace of single shot time domain transmission data for each parameter combination. The result is shown in \cref{fig:phase_diagram}(c) and (d), which shows the three regions traced out on the $\Delta-\eta$ parameter plane and the measured dwell time at half filling (dashed lines), respectively. 

The phase boundaries (points) are obtained from measured time-domain single shot telegraph transmission data as shown in \cref{fig:phase_diagram} (e)-(h) for the range of $\eta$ at $\Delta=0$ indicated in panel (a) with a double arrow. Here we define a threshold (dashed line) as described earlier and count if a single phase switching attempt was successful to cross this threshold within the measurement time. If the answer is yes the corresponding $\eta$ and $\delta$ value pair is assigned to the bistable region boundary. If the answer is no, depending on the measured value (below or above threshold) the parameter combination is assigned to the dim or bright phase. The parameter region where multiple crossings occur is assigned to the bistable region. For each detuning the detection bandwidth and total measurement time has been optimized to be able to determine a sharp phase boundary and to be able to resolve the dwell time over 5 orders of magnitude, as shown in Fig.~\ref{fig:phase_diagram}(d).

The raw data in \cref{fig:phase_diagram} (e)-(h) reveals an interesting difference between partial phase switching attempts from the dim state, which are quite frequent; and from the bright state, which are rather rare and typically of smaller amplitude (not visible in this data). This asymmetry is not observed in the simulated quantum jump trajectories as shown e.g. in \cref{fig:trajectoriesGammaPhi}, and its origin is not clear. However, these data point at an additional stabilization mechanism of the bright phase that might also contribute to the stronger than expected scaling towards the thermodynamic limit. 

Comparing the theoretical and experimental phase diagrams (a-c), one can see that the semiclassical 2-level model (similar to the neoclassical model) fails to capture the essential features of the experiment. The agreement with the 3-level case is qualitatively much better especially in case of the bright phase boundary. However, the shape of the lower limiting curve is not correctly captured by this theory. In fact, the 3-level semiclassical theory is unable to  reproduce the well pronounced dim phase region around resonance that we observe experimentally. In general it reproduces the overall resonance-like dependence on the detuning, and also the asymmetry with respect to the $\Delta=0$ line, but no quantitative agreement in terms of shape or absolute values could be obtained. 
This comparison with the semiclassical theory underlines that the well-resolved spectrum of the strongly coupled transmon-resonator system with more than two transmon levels plays an essential role in our experiment.



\section{Discussion, conclusions and outlook}
\label{sec:concl}

It is important to distinguish the presented PBB phase transition and scaling from similar related phenomena. The oldest known such effect is optical bistability, dispersive or absorptive, that is itself a first-order DPT \cite{drummond1980quantum,Casteels2016,Casteels2017}. In the case of the PBB, we are not in the dispersive regime however. The driving is close to or on resonance with the bare resonances of the resonator and the transmon, and the absorption of the latter does not play an essential role either \cite{Vukics2019}. 
Another related model is the Duffing oscillator that appears in a circuit QED context as a Kerr-nonlinear mode (the transmon) interacting with a linear mode (the resonator) \cite{Rebic2009,peano2010dynamical}. Parametric driving can lead to critical behavior \cite{Wustmann2019} and driven nonlinear inductors have exhibited slow classical switching events triggered by low frequency thermal fluctuations on the order of seconds \cite{Muppalla2018}.
Long bit flip times up to 100 seconds have also been observed in a two-photon dissipative oscillator that is characterized by symmetry breaking of the intra-cavity field phase, but this system does not exhibit a bistability in the photon number \cite{Berdou2022}. 
In this respect, the transmon, even with many levels considered, is algebraically very different from a nonlinear oscillator when it comes to jump operators because these are not bosonic. This makes an essential difference, as verified by our quantum simulations, where it is possible to try the consequences of different algebras. 
Our simulations clearly rule out the Duffing oscillator model, which cannot reproduce the phenomenology of the experiment since its bistable behaviour reminiscent of dispersive optical bistability occurs for different parameters and does not exhibit the same scaling towards the thermodynamic limit.
Finally, with respect to other recently-discovered QPTs and DPTs in the Jaynes-Cummings or Rabi models \cite{Hwang2015,Hwang2016,Larson2017Superradiant,Hwang2018}, where thermodynamic limits can also be defined in an abstract way, the difference of PBB as first-order DPT is that the thermodynamic limit is a strong-coupling limit. The well-resolved discrete spectrum of an interacting bipartite quantum system is essential for the effect.

In this paper, we have experimentally followed the finite-size scaling towards the $g/\kappa\to\infty$ thermodynamic limit with a characteristic time scale ranging over nearly seven orders of magnitude. Just like with a finite-size (non-macroscopic) sample of water, at $\SI{0}{\celsius}$ there is a contest of several meta- and even unstable states instead of true phases of liquid and ice \cite{thiele2019first}, in the PBB bistability for any finite value of $g$, there is a contest of non-macroscopically distinct dim and bright states. We have experimentally determined the scaling exponent of the bistable switching timescales of $\sim5.4\pm0.2$ as well as the finite-size scaling exponent of the intra-cavity photon number of the bright state of $\sim0.95\pm0.05$. We have also experimentally determined the phase diagram of the PBB phase transition and found that the characteristic dwell times drop by orders of magnitude for finite drive detunings.

We have compared these experimental results with large scale quantum simulations based on the QJMC method considering different numbers of transmon levels (3, 5 and 7) and different dephasing models of higher-lying levels. This comparison indicates that transmon levels up to at least 7 play an important role in the dynamics, as does the dephasing, since simulations with dephasing, e.g. due to flux noise, have shown better correspondence to the experimental data. Similarly, the comparison with semiclassical phase diagrams have also underlined the important role of higher transmon levels. Even though the full quantum simulations reproduce the observed trends correctly, there are significant differences from the experiment in the measured dwell times, which might indicate the presence of further stabilization mechanisms - in particular in the bright attractor - and calls for a better methods to model such strongly coupled multi-level systems. 

Even though the computational resources were substantial, the fully quantum numerical simulations were only suitable to model the three lowest coupling strengths $g/\kappa$ investigated. This highlights the need for powerful quantum simulators even in the case of comparably simple circuits and in particular to explain how macroscopic phases can be stabilized by individual quantum systems. It is in fact quite surprising that a single transmon qubit can switch back from the bright state - characterized by up to $10^4$ intra-cavity photons - all the way to the dim state and stabilize the empty cavity for seconds in the presence of the continuous large amplitude coherent input field - in particular given its limited potential confinement \cite{Shillito2022}. In the future, a fully confined qubit \cite{Hassani2022} with higher power handling, or larger anharmonicity and superconducting cavities with lower loss could help to explore even more macroscopic phases pushing the characteristic switching timescales from seconds to days.


Besides its fundamental interest as a quantum-classical phase transition, the PBB bistability also promises a few applications. Since single quantum jumps were shown to trigger the switching from the (nonclassical) dim state to the (closely classical) bright state \cite{Vukics2019}, our system may be considered as a quantum-jump amplifier, where ultimately a macroscopic microwave device (outside the fridge) is getting switched by microscopic quantum events (inside the fridge). An interesting prospect is controlling the switching behavior, that can be envisaged either in a parametric way, but preferably with another strongly coupled quantum system. In the latter case the bistability could act as a quantum readout device with high signal-to-noise ratio. The capability of preparing the system on the verge of a phase switching event could therefore make it applicable in quantum metrology and sensing based on microwave photon counting \cite{Guha2009}, a new paradigm for the application of first-order DPTs \cite{Lorenzo2017,raghunandan2018high,Yang2019,Heugel2019,DiCandia2021}.


The data and code used to produce the figures in this manuscript will be made available on the Zenodo repository.

\section*{Acknowledgments}
This work has received funding from the  Austrian Science Fund (FWF) through BeyondC (F7105) and the European Union's Horizon 2020 research and innovation program under grant agreement No 862644 (FETopen QUARTET). 
A.V. acknowledges support from the National Research, Development and Innovation Office of Hungary (NKFIH) within the Quantum Information National Laboratory of Hungary. The authors thank the MIBA workshop and the ISTA nanofabrication facility for technical support. We are grateful to the ELKH Cloud (\url{http://science-cloud.hu}) for providing us with the suitable computational infrastructure for the simulations. 

\appendix
 
\section{The full quantum model}
\label{app:fullQuantum}

\begin{figure*}[t]
    \centering
    \includegraphics[width=0.9\linewidth]{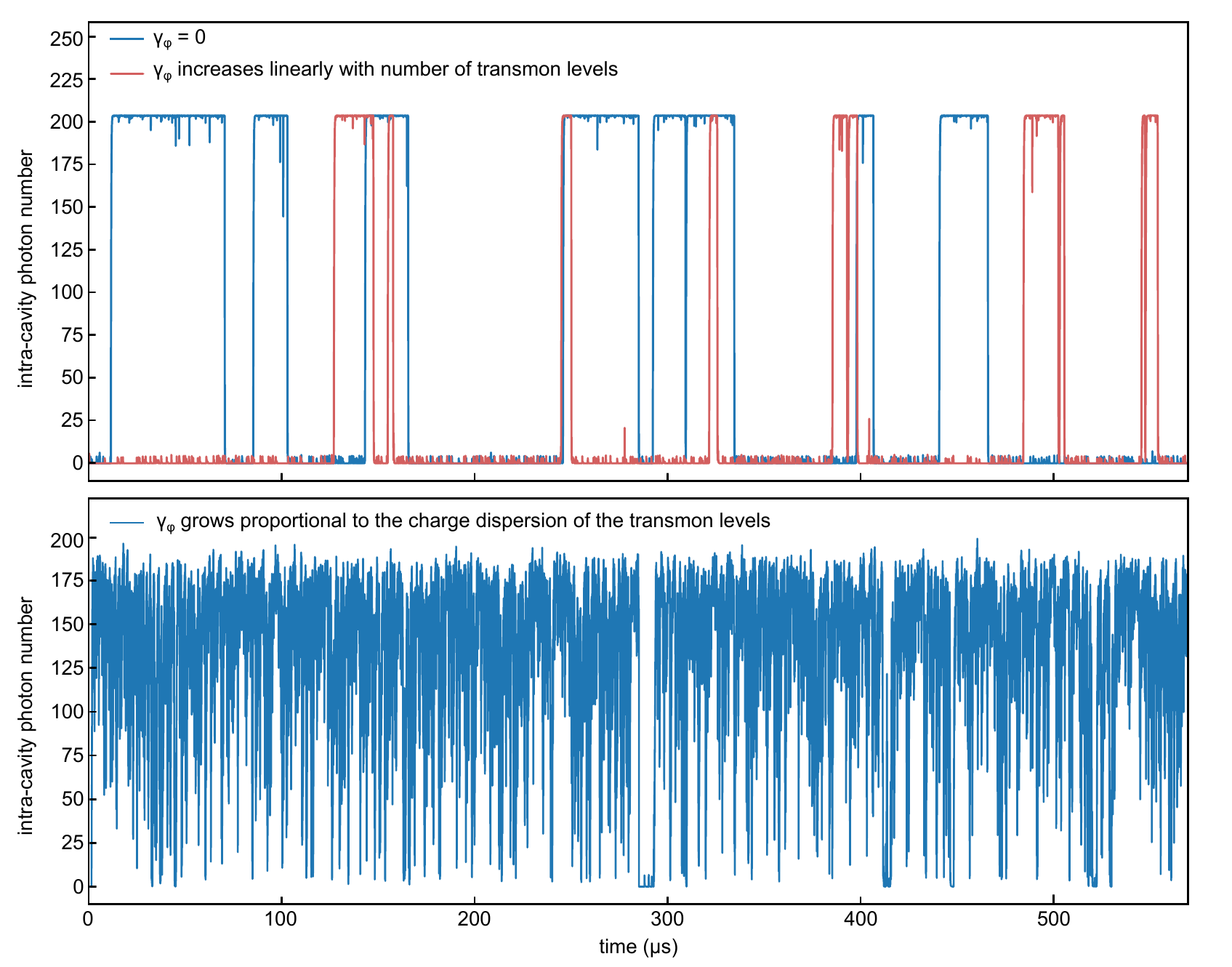}
    \caption{Simulated example trajectories for the three possibilities of modeling dephasing of higher transmon levels for otherwise identical parameters (five transmon levels, $\gamma_1=0$, $\gamma_{\phi,1}/(2\pi)=50\,$kHz, $\eta/(2\pi)=100\,$MHz, $\kappa/(2\pi)=14.3\,$MHz): no dephasing (blue), flux noise model (red) and charge dispersion model (blue, bottom).
The latter case leads to qualitatively incompatible results with very large noise levels and only partially stabilised attractors.}
    \label{fig:trajectoriesGammaPhi}
\end{figure*}

The basic Hamiltonian for a multi-level system interacting with a driven mode reads ($\hbar=1$):
\begin{equation}
\begin{aligned}
H=\sum_u\qty\Big[h_u\ket u\bra u + i\qty\big(g_{u+1}\,a\ket {u+1}\bra u-\text{h.c.})] \\
 + \omega_R\,a^{\dagger} a + i\,\qty(\eta\,e^{-i\omega t}\,a^{\dagger} - \text{h.c.}).
\end{aligned}
\end{equation}
Here, $u$ indexes the transmon levels, and we assume that only transitions between adjacent levels couple to the mode, with coupling coefficient $g_{u+1}$. The $h_u$s are the bare transmon energies, $\omega_R$ is the bare mode frequency, and $\eta$ and $\omega$ are the drive strength and frequency, respectively.

The Hamiltonian is written in the rotating-wave approximation. This is justified as long as the coupling strength does not reach the Bloch-Siegert regime of ultrastrong coupling, meaning $10\,g_{1}\lesssim h,\omega_R$ \cite{RevModPhys.91.025005}, which holds for our system. For the coupling coefficients we use the standard relation for transmons
\begin{equation}
g_{u+1}=\sqrt{u+1}\,g_{1}.
\end{equation}
For a comprehensive theory of the transmon cf. \cite{WallraffRMP,Schreier2008}.

Transforming to the frame rotating with $\omega$, we obtain a time-independent Hamiltonian with $\Delta\equiv\omega-\omega_R$:

\begin{equation}
\begin{aligned}
H=\sum_u\qty\Big[(h_u-u\,\omega)\ket u\bra u+i\qty\big(g_{u+1} a \ket {u+1}\bra u-\text{h.c.})] \\
 - \Delta a^\dagger a + i \qty(\eta a^\dagger-\text{h.c.}).
\end{aligned}
\end{equation}

Here, putting $h_0=0$, and assuming the $0-1$ transition resonant with the mode ($h_1=\omega_R$), we obtain a simple form for the bare transmon Hamiltonian, which we list for the first 3 levels:
\begin{equation}
\begin{aligned}
H_\text{transmon}=-\Delta \ket{1}\bra{1}-\qty(2\Delta-\Delta_\text{an}) \ket{2}\bra{2}\\
 +\text{contribution of higher levels},
\end{aligned}
\end{equation}
where $\Delta_\text{an}\equiv h_2-2 h_1$ is the anharmonicity of the third level, which is related to the charging energy.

Let us turn to dissipation, which we describe with the Liouvillian
\begin{equation}
\begin{aligned}
\mathcal{L}\rho=\sum_i\qty(L_i \rho L_i^\dagger-\frac12\qty{L_i^\dagger L_i \rho}) \\
\equiv\qty(\mathcal{L}_\text{mode}+\mathcal{L}_\text{relax}+\mathcal{L}_\text{dephase}) \rho
\end{aligned}
\end{equation}
with the following three dissipative channels:
\vspace{0.2cm}\\
\textbf{Resonator decay, $\mathcal{L}_\text{mode}$}
This is described by the jump operators $L_-=\sqrt{2\,\qty(n_\text{th}+1)\,\kappa}\,a$ and $L_+=\sqrt{2\,n_\text{th}\,\kappa}\,a^\dagger$. Here $n_\text{th}$ is the number of thermal photons, which can be neglected in our system, so the second kind of quantum jumps (absorption of thermal photons) does not exist.
\vspace{0.2cm}\\
\textbf{Energy relaxation of the transmon, $\mathcal{L}_\text{relax}$}
In analogy with the coupling to the resonator mode, we assume that this occurs only as transitions between adjacent levels. It is described by the jump operators $L_{u+1\to u}=\sqrt{\gamma_{u+1\to u}}\,\ket u \bra{u+1}$. In the simulation, we take $\gamma_{u+1\to u}$ equal for all levels, and we identify it with $\gamma_1$ in cQED.
\vspace{0.2cm}\\
\textbf{Dephasing of the transmon, $\mathcal{L}_\text{dephase}$}
This is also defined separately for all transmon levels, and its jump operator for level $v$ is
\begin{multline}
L_{\phi,v}=\sqrt{\gamma_{\phi,v}}\qty(\sum_{u\neq v}\ket u\bra u-\ket v\bra v)\\=\sqrt{\gamma_{\phi,v}}\qty\big(\mathbf{1}-2\ket v\bra v),
\end{multline}
so it simply flips the phase of level $v$ by $\pi$. Modeling the behavior of the dephasing for different transmon  levels is nontrivial. We consider three possibilities:
\begin{enumerate}
    \item $\gamma_{\phi,v}=0$ for all $v$. This is only to get a theoretical baseline of dephasing-free behavior.
    \item Linear growth as $\gamma_{\phi,v}=v\,\gamma_{\phi}/8$ following the above convention, as expected for flux noise due to the higher flux gradient of higher levels. Here the dephasing of the first qubit transition $\gamma_{\phi}/2\pi=\SI{50}{\kilo\hertz}$ is taken from the measured vacuum Rabi linewidths and the independently measured cavity linewidth $\kappa$.  
    \item Dephasing proportional to the charge dispersion of the transmon levels \cite{Bishop2009}.
\end{enumerate}
Example trajectories for the three possibilities are displayed in \cref{fig:trajectoriesGammaPhi}. It is apparent that model 3 leads to very noisy trajectories that do not reproduce qualitatively the experimentally observed behavior of stabilized attractors. Therefore, we omitted this possibility from the quantitative comparison presented in the main text.

\begin{figure*}[t]
    \centering
    \includegraphics[width=0.7\linewidth]{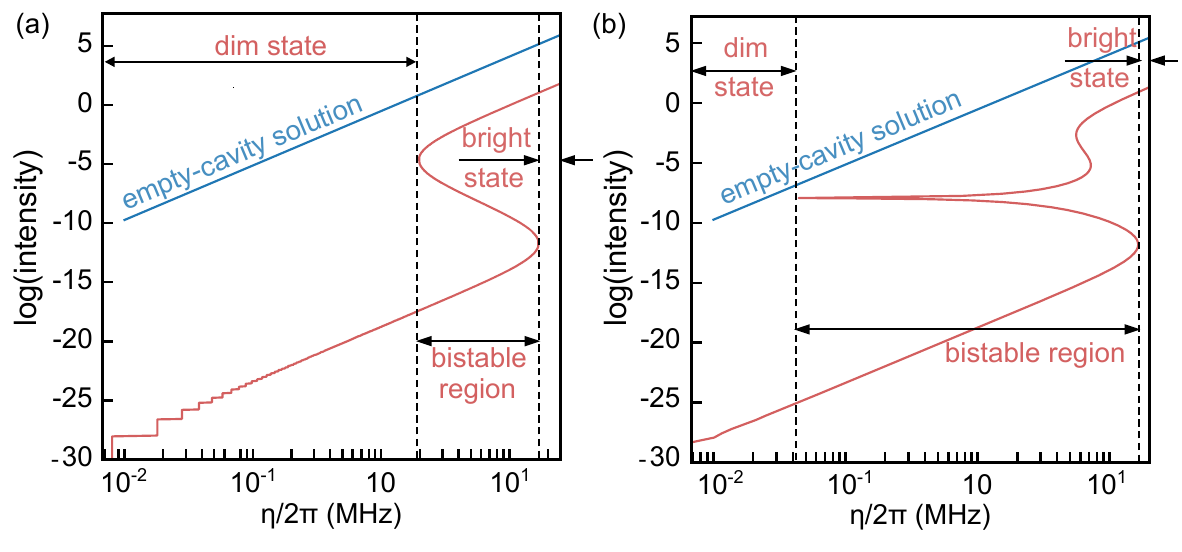}
    \caption{Typical bistability curves obtained from Maxwell-Bloch equations for the intra-cavity intensity as a function of the drive strength for $g/\kappa\approx132$ and detuning $\Delta/(2\pi)=-\SI{10}{\mega\hertz}$ for in case of 2 and 3 transmon levels in panels \textbf{(a)} and \textbf{(b)}, respectively ($\gamma_1/(2\pi)=1\,$kHz, $\gamma_{\phi,1}/(2\pi)=50\,$kHz). 
    The bistable region is characterized by a threefold (for certain parameters even fivefold) solution but one (two) that show a decreasing intensity with increasing drive strength are unphysical. 
    The empty-cavity solution (blue line) is shown for zero detuning. 
    }
    \label{fig:bistability}
\end{figure*}

In the simulation, for each physical parameter set, several trajectories are run with different random number generator seeds. Relying on the assumption of ergodicity, these trajectories are concatenated for a single long trajectory for each parameter set, which is then used for dwell-time statistics. Since each trajectory is started from the ground state, this method has a bias toward the dim state (breaching of ergodicity), which is the stronger, the larger the dwell time with respect to the simulation time.

The full quantum simulations were implemented within the C++QED simulation framework (\url{http://github.com/vukics/cppqed}), and took about a year on a 64-core virtual cluster defined within an OpenStack Cloud environment (\url{http://science-cloud.hu/}).

\section{The Maxwell-Bloch equations and their solution for the intensity}
\label{app:semiclassical}

From the master equation $\dot\rho=\qty[H,\rho]/(i\hbar)+\mathcal{L}\rho$ we can derive equations for the expectation values of the operators $a$ and $\sigma_{uv}=\ket u\bra v$. In the case of a two-level system, this simply reproduces the Maxwell-Bloch equations, with the added complication of the qubit dephasing. Here, we list the equations for a three-level transmon with states $|g\rangle$, $|e\rangle$ and $|f\rangle$, which still leads to an algebraically tractable scheme. In this case, 6 equations are needed for a complete system:
\begin{subequations}
\label{eq:semiclassical}
\begin{align}
\label{eq:alpha}
    \dot\alpha=&(i\Delta-\kappa)\alpha+\eta-g_1\,s_{ge}-g_2\,s_{ef}\\
\label{eq:sge}
    \dot s_{ge}=&\qty(i\Delta-\gamma_1)\,s_{ge}-g_1\,\qty(s_{ee}-s_{gg})\,\alpha-g_2\,s_{gf}\,\alpha^*\\
\label{eq:sgg}
    \dot s_{gg}=&\gamma_1\,s_{ee}-2g_1\,\real\qty{\alpha^*\,s_{ge}}\\
\label{eq:sef}
    \dot s_{ef}=&\qty(i\qty[\Delta-\Delta_f]-\qty[\gamma_1+4\qty(\gamma_{\phi,1}+\gamma_{\phi,2})])\,s_{ef}\nonumber\\&+g_2\,\qty(s_{ee}-s_{ff})\alpha+g_1\,\alpha^*\,s_{gf}
\\
\label{eq:see}
    \dot s_{ee}=&2\,g_1\,\real\qty{\alpha^*\,s_{ge}}-2\,g_2\,\real\qty{\alpha^*\,s_{ef}}-\gamma_1\,s_{ee}+\gamma_1\,s_{ff}\\
\label{eq:sgf}
    \dot s_{gf}=&\qty(i\qty[2\Delta-\Delta_f]-\qty[\gamma_1+4\gamma_{\phi,2}])\,s_{gf}\nonumber\\&-g_1\,\alpha\,s_{ef}+g_2\,\alpha\,s_{ge}
\end{align}
\end{subequations}
Here $\alpha=\expval{a}$, and $s_{uv}=\expval{\sigma_{uv}}$. The system is completed with the completeness relation $s_{gg}+s_{ee}+s_{ff}=1$. We are interested in the steady state, which can be obtained by zeroing the left hand side of the equations, that leads to an inhomogeneous nonlinear set of equations.

We do not need to solve the full set of equations. Instead, we can obtain a single implicit equation for only the intensity $\abs{\alpha}^2$ as follows. First, we define the complex dispersive shift
\begin{equation}
\label{eq:CDS}
\Sigma\qty(\abs{\alpha}^2)=-\frac{g_1\,s_{ge}+g_2\,s_{ef}}{\alpha},
\end{equation}
then, from \cref{eq:alpha} in steady-state we express $\alpha$ explicitly. As we will show below, $\Sigma$ depends only on powers of $\abs{\alpha}^2$, and not on other combinations of $\alpha$ and $\alpha^*$. Therefore, the equation for the intensity can be written as
\begin{equation}
\label{eq:implicitIntensityEquation}
\abs{\alpha}^2=\frac{\abs{\eta}^2}{\abs{\Sigma\qty(\abs{\alpha}^2)+(i\Delta-\kappa)}^2}
\end{equation}

What we have to show for the validity of \cref{eq:implicitIntensityEquation} is that the solutions of $s_{ge}$ and $s_{ef}$ have the form of an $\abs{\alpha}^2$-dependent expression multiplied by $\alpha$. The polarizations can be expressed as functions of the populations multiplied with $\alpha$ from \cref{eq:sge,eq:sef,eq:sgf} in steady state. When these solutions are substituted into the steady-state population equations \labelcref{eq:sgg,eq:see}, the factor $\alpha$ in the solutions together with $\alpha^*$ in those equations give an $\abs{\alpha}^2$. Hence, the populations can be expressed from these equations as functions only of the intensity, and when these are substituted back into the solutions of the polarizations, we obtain the necessary form for these latter.

A typical solution of \cref{eq:implicitIntensityEquation} exhibiting bistability is displayed in \cref{fig:bistability}. The semiclassical theory is inferior to the full quantum-trajectory solution described in the main text and \cref{app:fullQuantum} in at least two respects:
\begin{enumerate}
    \item Dealing with (possibly multi-valued) steady state solutions, it does not provide information on timescales.
    \item Whereas the set of three complex polarizations and two populations in \cref{eq:semiclassical} give a complete picture of the transmon in itself, the mode is represented only by a single amplitude. This means that the theory cannot account for nonclassical states of the mode and transmon-mode entanglement.
\end{enumerate}

\bibliography{PBB_prx_Bibliography}

\end{document}